\documentclass[
    prl,
    10pt,
    aps,
    twocolumn,
    floatfix,
    showpacs,
    noshowpacs,
    longbibliography,
    superscriptaddress, 
]{revtex4-1}
\usepackage[T1]{fontenc}
\usepackage{amsmath}
\usepackage{amssymb}
\usepackage{epsfig}
\usepackage{multirow}
\usepackage{bm}
\usepackage{color}
\usepackage{hyperref}
\usepackage{url}
\usepackage{graphicx}

\newcommand{\abs}[1]{\lvert #1 \rvert}

\newcommand{\beq}{\begin{equation}}
\newcommand{\eeq}{\end{equation}}
\newcommand{\bea}{\begin{eqnarray}}
\newcommand{\eea}{\end{eqnarray}}

\newcommand{\eF}{\varepsilon_{F}}

\newcommand{\kF}{k_F}

\newcommand{\din}{d^{}_{i}}
\newcommand{\dout}{d^{}_{f}}
\newcommand{\Eflow}{E_j}

\newcommand{\dx}{\mathrm{d}x}
\providecommand{\exclude}[1]{}

\newcommand{\vbr}{{\bf r}}
\newcommand{\orcidicon}[1]{\href{https://orcid.org/#1}{\includegraphics[height=\fontcharht\font`\B]{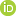}}}

\begin{document}

\title{Dissipative Dynamics of Quantum Vortices in Fermionic Superfluid}

\author{Andrea Barresi\,\orcidicon{0000-0002-9769-8702}}\email{andrea.barresi.dokt@pw.edu.pl}
\affiliation{Faculty of Physics, Warsaw University of Technology, Ulica Koszykowa 75, 00-662 Warsaw, Poland}

\author{Antoine Boulet\,\orcidicon{0000-0003-3839-6090 }}\email{antoine.boulet@pw.edu.pl}
\affiliation{Faculty of Physics, Warsaw University of Technology, Ulica Koszykowa 75, 00-662 Warsaw, Poland}

\author{Piotr Magierski\,\orcidicon{0000-0001-8769-5017}}\email{piotr.magierski@pw.edu.pl}
\affiliation{Faculty of Physics, Warsaw University of Technology, Ulica Koszykowa 75, 00-662 Warsaw, Poland}
\affiliation{Department of Physics, University of Washington, Seattle, Washington 98195--1560, USA}

\author{Gabriel Wlaz\l{}owski\,\orcidicon{0000-0002-7726-5328}}\email{gabriel.wlazlowski@pw.edu.pl}
\affiliation{Faculty of Physics, Warsaw University of Technology, Ulica Koszykowa 75, 00-662 Warsaw, Poland}
\affiliation{Department of Physics, University of Washington, Seattle, Washington 98195--1560, USA}

\begin{abstract}
In a recent article, Kwon \textit{et al.} [Nature (London) {\bf 600}, 64 (2021)] revealed nonuniversal dissipative dynamics of quantum vortices in a fermionic superfluid. The enhancement of the dissipative process is pronounced for the Bardeen-Cooper-Schrieffer interaction regime, and it was suggested that the effect is due to the presence of quasiparticles localized inside the vortex core. 
We test this hypothesis through numerical simulations with time-dependent density-functional theory: a fully microscopic framework with fermionic degrees of freedom. The results of fully microscopic calculations expose the impact of the vortex-bound states on dissipative dynamics in a fermionic superfluid. 
Their contribution is too weak to explain the experimental measurements, and we identify that thermal effects, 
giving rise to mutual friction between superfluid and the normal component, dominate the observed dynamics.
\end{abstract}

\maketitle

\paragraph{Introduction.---}Quantum simulators in the form of ultracold atoms with fine-tuned interactions offer a versatile platform for studying many-body phenomena in quantum systems. In particular, an emergent phenomenon of superfluidity is the subject of extensive studies. 
Presently, the effort has been shifted toward the investigation of mechanisms that lead to energy dissipation, although the underlying system has formally vanishing viscosity coefficients. 
Recent experiments at LENS (Florence, Italy) highlighted astonishing dissipative processes during the scattering of quantum vortices~\cite{Kwon:2021a}. In this experiment, the relative distance change between quantum vortices during the collision was used as a probe that quantifies the collective energy losses. 
The conclusions are unequivocal: the dissipation changes as we change the nature of the underlying superfluid from bosonic to fermionic and is significantly enhanced for the latter case.
Sensitivity of the superfluid dynamics with respect to the regime has also been tested in measurements of critical velocity~\cite{PhysRevLett.99.070402,PhysRevLett.114.095301,PhysRevLett.121.225301} or behavior of an atomic Josephson junction ~\cite{PhysRevLett.120.025302,PhysRevLett.124.045301}.

Experiment~\cite{Kwon:2021a} has been conducted with a fermionic isotope of ${}^{6}$Li, cooled down to superfluid phase. To characterize the interaction regime, it is convenient to introduce the dimensionless quantity $a_s k_F$, where $a_s$ is the $s$-wave scattering length and $k_F=(3\pi^2n)^{1/3}$ is the Fermi wave vector corresponding to the density $n$. 
The Bose-Einstein condensate (BEC) regime corresponds to positive and small values of $a_s k_F \rightarrow 0^+$, where bound states (dimers) are created that behave effectively as bosons. The amount of measured dissipation is relatively small for the BEC regime, and the zero-temperature Gross-Pitaevskii equation (GPE) is able to explain the measurements~\cite{Kwon:2021a} successfully. The GPE points to the emission of phonons (sound) as the primary dissipation mechanism. On the other side, when $a_s k_F \rightarrow 0^-$, fermions with opposite spins form quantum correlations in the form of Cooper pairs. It corresponds to the Bardeen-Cooper-Schrieffer (BCS) coupling regime. 
In this regime, a significant enhancement of the collective energy dissipation is observed. It is speculated that an additional dissipation mechanism activates in this regime, genuinely related to the fermionic nature of the system. The enhanced dissipation is also present for strongly interacting case, called unitary Fermi gas (UFG), where $a_s k_F \rightarrow \pm \infty$, however, not as strong as in the BCS limit. This Letter aims to provide microscopic insight into the dissipative processes for fermionic systems with strong (UFG) and weak (BCS) interactions. 

In a pioneering work~\cite{Silaev:2012a}, a universal dissipation mechanism induced by the motion of the topological defects, and present only in fermionic superfluids, was proposed. The mechanism is due to the presence of the internal structure of quantum vortices: in the Fermi system, the vortices host localized Andreev states, implying that their cores are filled with a gas of quasiparticles~\cite{PhysRevLett.94.140401,PhysRevLett.96.090403,Machida2007-lf,PhysRevA.106.033322,PhysRevLett.80.2921,doi:10.1143/JPSJ.67.3368}.
When they move with an acceleration, these Andreev quasiparticles can be excited and eventually converted into delocalized states.
Occupation of the Andreev states is affected, which in Ref.~\cite{Silaev:2012a} is interpreted as an increase in the vortex core's effective temperature. The internal structure of quantum vortices is not considered in GPE-like approaches or phenomenological approaches like the vortex filament model, and the fact that they failed in explaining observations of~\cite{Kwon:2021a}, for UFG and BCS regimes, directs to speculation that the mechanism as proposed by Silaev can be responsible for the observed discrepancy. 
The same mechanism is expected to be the main source that differentiates energy dissipations between ${}^{3}$He-B (fermionic) and ${}^{4}$He (bosonic) superfluids~\cite{Eltsov2014,Autti2020}. 
Its microscopic understanding is important in the context of all types of Fermi superfluids~\cite{Volovik:bookHe3,Volovik2009,Graber2017,Pethick2002} and is still missing.

This work provides a large-scale simulation to study the scattering of vortices, aiming to expose the microscopic origin of dissipation observed in the experiment~\cite{Kwon:2021a}. Our approach is based on time-dependent density-functional theory (TDDFT). The theory utilizes explicitly fermionic quasiparticles as degrees of freedom, and thus effects due to Andreev states are naturally incorporated. 
Nowadays, energy functionals for superfluid Fermi gases have reached a high level of maturity, allowing systematic and accurate studies of the systems that facilitate comparison with experiments~\cite{Boulet:2022a}.
We study the cases for which the TDDFT method has been extensively validated: UFG regime ($|a_s k_F|=\infty$), where the so-called superfluid local density approximation (SLDA) functional proved to be accurate~\cite{Bulgac2012,Wlazlowski2018,PhysRevA.105.013304,PhysRevA.104.053322,PhysRevLett.102.085302,Bulgac2011,PhysRevLett.108.150401,PhysRevA.91.031602,PhysRevA.100.033613,PhysRevA.103.L051302,PhysRevA.104.033304,PhysRevLett.112.025301}, and BCS regime $|a_s k_F| \lesssim 1$, where the Bogoliubov--de Gennes (BdG) functional is trustable. We first revisit the static properties of fermionic quantum vortices and identify energy scales that are important for dynamical simulation. 

\begin{figure*}[t]
    \centering
    \includegraphics[scale=0.85]{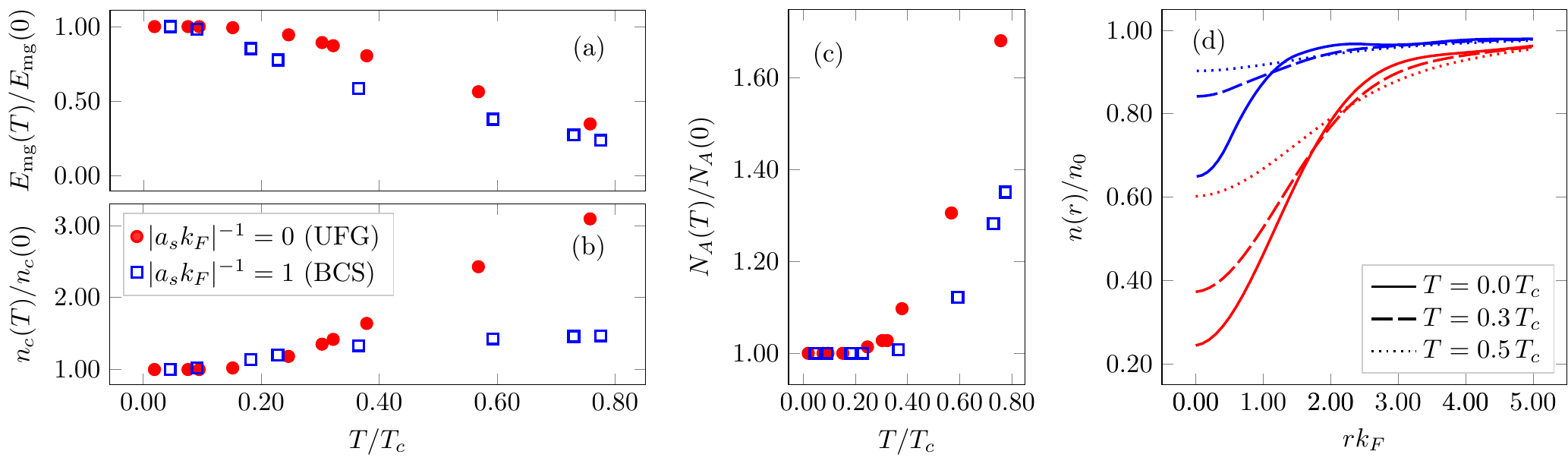}
    \caption{(a) Minigap energy and (b) density at the center of the vortex core $n_c$ as a function of the dimensionless temperature of single vortex
    at unitarity (red filled circles) and in BCS regime (blue open squares). For convenience, the quantities are displayed according to their values obtained at zero temperature.  
In (c), we show temperature evolution of the number of Andreev states ($E_n\le 0.9|\Delta|$) residing in the vortex. Vortex density profiles as a function of the distance from the core in UFG (red) and BCS (blue) at $T = 0$ (solid line), $T = 0.3T_c$ (dashed line), and $T = 0.5T_c$ (dotted line) are shown in (d). \label{fig:E_mg-and-rho_c-T} }
\end{figure*}

\paragraph{Structure and typical scales of quantum vortex. ---}
The static variant of density-functional theory (DFT) we apply here is formally equivalent to the mean-field  Bogoliubov--de Gennes equations 
\begin{equation}
\mathcal{H}(n,\nu)
\begin{pmatrix}
u_n({\bf r})\\
v_n({\bf r})\\
\end{pmatrix}=E_n \begin{pmatrix}
u_n({\bf r})\\
v_n({\bf r})\\
\end{pmatrix}
\label{eqn:stbdg}
\end{equation}
for Bogoliubov amplitudes $(u_n({\bf r}),v_n({\bf r}))^{T}$ that define normal $n$ and anomalous $\nu$ densities 
\begin{subequations} \label{eq:bdg-densities}
	\begin{align}
		n(\vbr)
		& = 2\sum_{E_n>0}\left(\abs{u_{n}(\vbr)}^2f^+_n + \abs{v_{n}(\vbr)}^2f^-_n\right), \label{eq:bdg-densities-n}
		\\
		\nu(\vbr)
		& = \sum_{E_n>0} (f^-_n - f^+_n)u_{n}(\vbr)v_{n}^{*}(\vbr).\label{eq:bdg-densities-nu}
	\end{align}
\end{subequations}
The Fermi-Dirac distribution, noted as $f_{n}^\pm = [1 + \exp \left ( \pm E_n/T \right )]^{-1}$, is included to model the temperature $T$ effects. We use the metric system, where $m=\hbar=k_B=1$. The Hamiltonian has generic form
\begin{equation}
\mathcal{H} = \begin{pmatrix}
-\frac{1}{2}\nabla^2+U({\bf r}) - \mu & \Delta({\bf r}) \\
\Delta^*({\bf r}) & \frac{1}{2}\nabla^2-U({\bf r}) + \mu \\
\end{pmatrix},\label{eqn:HBdG}
\end{equation}
where mean and pairing fields are computed as appropriate functional derivatives of the energy functional $\mathcal{E}$, namely $U=\frac{\delta\mathcal{E}}{\delta n}$ and $\Delta=-\frac{\delta\mathcal{E}}{\delta \nu^{*}}$. Explicit forms of these fields depend on the interaction regime. For the BCS regime they are $U^{(\textrm{BCS})}=0$ and $\Delta^{(\textrm{BCS})}=-g \nu$ with $g\sim 4\pi a_s$. 
With these definitions, the method becomes identical with celebrated BCS theory when applied to a uniform system. 
For UFG, we use a functional known as SLDA~\cite{Bulgac2007}, which gives $U^{(\textrm{UFG})}=\frac{\beta(3\pi^2 n)^{2/3}}{2}-\frac{|\Delta|^2}{3\gamma n^{2/3}}$ and $\Delta^{(\textrm{UFG})}=-\frac{\gamma}{n^{1/3}} \nu$. This form of the fields assures us that the theory is scale invariant. 
Coupling constants $\beta$ and $\gamma$ are adjusted to ensure the correct energy value $E/N=\frac{3}{5}\xi_0\eF$ with Bertsch parameter $\xi_0\approx0.4$ and energy gap $\Delta/\eF\approx 0.5$, when used for the uniform system. Here, $\eF=k_F^2/2$ stands for the Fermi energy. 
The total particle number $N$ is controlled by the chemical potential $\mu$. In the presence of the external trapping potential one needs to redefine the mean field $U({\bf r})\rightarrow U({\bf r})+V_{\textrm{ext}}({\bf r})$. The coupling constants that define the pairing field ($g$ and $\gamma$) need to be renormalized in order to remove formal divergence of anomalous density as given by Eq.~(\ref{eq:bdg-densities-nu}). 
It is done by introducing energy cutoff $E_c$ at which the sum is truncated $\sum_{E_n>0}\rightarrow \sum_{0<E_n<E_c}$, see~\cite{Boulet:2022a} for a more detailed discussion. 

The minigap energy $E_{\textrm{mg}}$ is a crucial quantity when discussing fermionic vortices~\cite{Volovik2009,PhysRevA.106.033322,PhysRevC.104.055801}. It is defined as the energy of the lowest Andreev state. 
In BCS approximation, at $T=0$, we have $E_{\textrm{mg}}\approx |\Delta|^{2}/2\eF$. This formula works reasonably well in the entire BCS-UFG range, taking  that $\Delta/\eF\approx\frac{8}{e^2}\exp(-\pi/2|a_s\kF|)$ for the BCS and $\Delta/\eF\approx 0.5$ for the UFG regimes~\cite{Boulet:2022a}.
The number of Andreev states (below the energy gap) scales as $N_A \sim |\Delta|/E_{\textrm{mg}}$, and clearly it increases exponentially as we move toward the deep BCS limit, see also~\cite{PhysRevB.99.134506}. Thus, the vortices in the BCS regime host more matter inside as compared to the UFG limit, compare vortex profiles presented in Fig.~\ref{fig:E_mg-and-rho_c-T}(d). 
This naturally suggests the increasing role of the vortex core structure on the dynamical properties as we move from UFG to BCS interaction regimes.

The experiment~\cite{Kwon:2021a} was conducted for temperature $T/T_c\approx 0.3$--$0.4$, where $T_c$ is the critical temperature of the superfluid-normal phase transition. 
We have checked sensitivity of the vortex solution with respect to the temperature effects for UFG ($|a_s\kF|^{-1} = 0$) and for BCS ($|a_s\kF|^{-1} = 1$) regimes. The results are presented in Fig.~\ref{fig:E_mg-and-rho_c-T}. For the strongly interacting unitary gas, the minigap energy $E_{\textrm{mg}}$, the vortex core density $n_c$, and number of Andreev states $N_A$ are almost independent of the temperature for $T \lesssim 0.2\,T_c$~\cite{TcNote}. Above it, the temperature dependence for the quantities is clearly visible.
The BCS regime case exhibits different behaviors of the static properties as compared to UFG. The minigap energy change is observed already for temperatures close to zero. The density at the center of the vortex core reaches approximately the bulk density value, already at $T \simeq 0.3\,T_c$. 
Clearly, for temperatures achieved in experiment, the vortex solution in UFG is affected by thermal effects, and in the case of the BCS regime, the thermal impact becomes significant. These aspects suggest that the zero-temperature formalism may fail to explain the results of experiment~\cite{Kwon:2021a}, and the most likely observed enhancement of the dissipation in BCS is of dual origin: due to mutual friction with the normal component and internal structure of quantum vortices. Dynamical calculations are needed to specify relative importance of these two. 

It is interesting to note that the matter density inside the vortex core increases with temperature, see Fig.~\ref{fig:E_mg-and-rho_c-T}(b) and~\cite{PhysRevB.87.214507}. It allows one to use the core density as a probe that measures the vortex's (local) temperature. 
Suppose the mechanism as proposed by Silaev~\cite{Silaev:2012a} is in action. In that case, we expect to see an increase in the core density after the vortex collision; according to the interpretation, the dissipative process heats up the vortex.  

\paragraph{Propagation and collision of vortices. ---}
The vortex dynamics is studied by means of TDDFT formalism. It is obtained from the static variant by replacing $u_n({\bf r})\rightarrow u_n({\bf r},t)$, and similarly for the $v_n$ component, and converting Eq.~(\ref{eqn:stbdg}) to time-dependent form by applying $E_n\rightarrow i\partial / \partial t$.  
It was already emphasized that the impact of temperature effects may be significant; thus time-dependent calculations should take these effects into account. While the DFT formalism can be rigorously extended to finite temperatures~\cite{PhysRev.137.A1441,PhysRevB.82.205120}, there is no such extension to the time-dependent problems. 
The simplest way is to assume that densities~(\ref{eq:bdg-densities}) acquire time dependence only through $\{u_n({\bf r},t),v_n({\bf r},t)\}$, while the Fermi-Dirac distribution function is kept to be frozen. This procedure is justified if the system stays close to the equilibrium all time during the dynamics, otherwise it constitutes an uncontrolled approximation.
A more refined approach would be to allow the distribution $f_{n}^\pm$ to evolve in time as well, for example, by coupling theory to the Boltzmann equation as it was done in the case of Bose system within the Zaremba-Nikuni-Griffin approach~\cite{Zaremba1999}. Practical realization of this concept for the Fermi system has not been demonstrated. An alternative approach of incorporating fluctuations and dissipation within TDDFT was proposed in~\cite{Bulgac2019}. 
Contrary to the mentioned extensions, the approach we applied does not introduce additional (phenomenological) parameters to the theory, which eventually one should treat as a fitting parameter. 

\begin{figure*}[t!]  
\includegraphics[width=\textwidth]{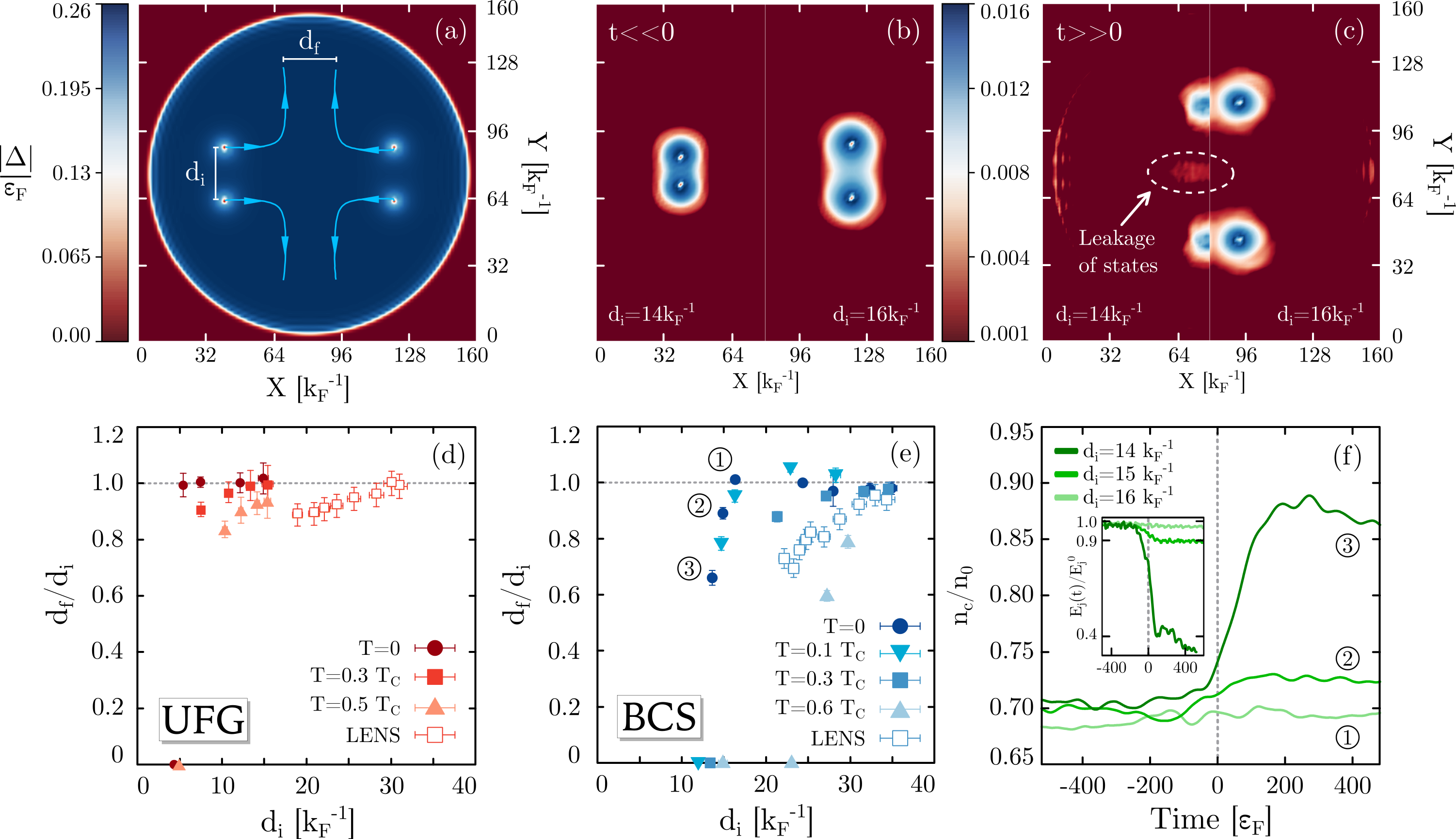}
    \caption{(a) Initial configuration (BCS regime with $a_s k_F=-1$) showing distribution of the order parameter $\Delta$. During the dynamics, vortices are moving along blue lines; see examples movies in Supplemental Material~\cite{SM}. Distance between vortices before and after collision is indicated by $\din$ and $\dout$, respectively. 
    (b), (c) Spatial distribution of density arising from the Andreev states only $n_A(\vbr)$, before ($t\ll0$) and after collision ($t\gg0$). Boxes are divided into half, corresponding to different initial distances of vortices in the BCS regime. 
    (d), (e) Relative decrease in distance between vortices in the case of two dipoles colliding head-on in UFG and BCS regimes at various temperatures. Error bars account for finite resolution of the computational lattice. For reference, we also provide experimental results of LENS~\cite{Kwon:2021a}. (f) Core density, normalized to the bulk density, as a function of time for zero-temperature BCS runs. Time $t=0$  indicates collision moment. 
    Inset: flow energy $\Eflow$ as a function of time, normalized to its initial value. Each energy line matches its color. Lines marked by numbers 1--3 correspond to points with the same labels as in (e). 
    }
    \label{fig:quad}
\end{figure*}

We consider head-on collisions of vortex dipoles: two vortices of opposite circulation that move parallel to each other, assuming that the intervortex distance is bigger than a threshold value for the pair annihilation. 
The calculations are executed by solving the time-dependent equations on a spatial lattice of size $100\times 100\times 16$, where in the $z$ direction we assume that the system is uniform. The lattice spacing was set to satisfy $\xi/\dx\approx 2.0$ in BCS and $\xi/\dx\approx 1.6$ in UFG regimes, where $\xi=\kF/\pi\Delta$ is the BCS coherence length, which assures reasonable representation of the Andreev states~\cite{PhysRevA.103.L051302}.  The system is trapped in a cylindrical external potential, similar to the experimental setup~\cite{Kwon:2021a}. The initial solution with four quantum vortices is obtained through the imprinting technique.
The number of particles  $N=\int n(\vbr)\,d^3\bm{r}$ is adjusted in such way to get $\kF\simeq1.6$ and $\kF\simeq0.8$ for BCS and UFG regimes respectively, where $\kF$ is defined through density in the trap center. The numerical setup is presented in Fig. \ref{fig:quad}(a); see also the Supplemental Material~\cite{SM} for details related to imprinting of vortices.
 
In Figs.~\ref{fig:quad}(d) and~\ref{fig:quad}(e), we present numerically obtained change in  the intervortex distance $\dout/\din$ due to the collisions. In the case of UFG, we find that at $T=0$ collisions are essentially elastic ($\dout/\din\simeq 1$), up to the annihilation threshold [Fig.~\ref{fig:quad}(d)].
We observe a decrease of the intervortex distance $\dout$ only if we increase the temperature up to $T/T_c\gtrsim0.3$, which matches the temperature required to induce changes in the vortex structure, see also Fig.~\ref{fig:E_mg-and-rho_c-T}. On the other hand, in the BCS regime, we find that already at $T=0$ dissipative dynamics emerge for cases close to the annihilation threshold [Fig.~\ref{fig:quad}(e)]. As expected, the dissipation as measured by the ratio $\dout/\din$ is further enhanced for the finite-temperature runs. For temperatures $T/T_c\gtrsim0.3$, we find that the suppression of $\dout/\din$ is mainly dominated by the thermal effects.  

To clarify the origin of the dissipative dynamics in the BCS regime at $T=0$, we have analyzed the vortex structure evolution during the process. In Fig.~\ref{fig:quad}(f), we present matter density in the vortex as a function of time. We see that, for cases where $\dout/\din<1$, the density increases due to the collision. It demonstrates that the process becomes sensitive to the vortex core structure. To visualize the process explicitly, in Figs.~\ref{fig:quad}(b) and~\ref{fig:quad}(c), we provide the evolution of density arising only from the Andreev states,
\begin{equation}
n_{A}(\vbr)
		 = 2\sum_{0<E_n<0.85\Delta}\left(\abs{u_{n}(\vbr)}^2f^+_n + \abs{v_{n}(\vbr)}^2f^-_n\right). 
\end{equation}
Before the collision, the density contracted from the in-gap states is entirely localized to the region where the topological defects are present, Fig.~\ref{fig:quad}(b). During the collision, their distribution is affected and some of these states become even delocalized, visible as leakage of density $n_{A}$ from the cores.
Effectively, the vortices emerge as being heated up after the collision.
The strength of this process is related to $\din$, which in turn is related to the acceleration of moving vortices: smaller $\din$ generates a trajectory of higher curvature and thus higher centripetal acceleration. 
The presence of the dissipative process is reflected also in a drop of the flow energy $\Eflow=\int \frac{\bm{j}^2}{2n}\,d^3{\bm{r}}$, shown  in  inset  of  Fig.~\ref{fig:quad}(f), with current computed as 
\begin{equation}
\bm{j}(\vbr) = 2\sum_{E_n>0} \bigg\{  \textrm{Im}[u^{*}_{n}(\vbr)
\nabla u_{n}(\vbr)]f^+_n 
- \textrm{Im}[v^{*}_{n}(\vbr)\nabla v_{n}(\vbr)]f^-_n
\bigg\}.
\end{equation}
The $\Eflow$ energy contains contributions from incompressible $\Eflow^{\textrm{(i)}}$ (vortices) and compressible $\Eflow^{\textrm{(c)}}$ (sound) modes~\cite{PhysRevLett.78.3896,Tsubota2017}, and conversion $\Eflow^{\textrm{(i)}}\rightarrow\Eflow^{\textrm{(c)}}$ is also detected during the collision. The observed suppression of $\dout/\din<1$ at zero temperature is mainly due to effects related to the core structure. They share similarities with the mechanism predicted by Silaev~\cite{Silaev:2012a}, which was derived based on quasi-classical arguments (vortex is approximated as a container that holds gas of quasiparticles). Here, we demonstrate importance of the vortex core structure on the dynamics from the perspective of the microscopic description, which extends beyond capabilities of the Gross-Pitaevskii approach, which is the only one that was used to study vortex collisions so far~\cite{Kwon:2021a,Yang2016,Yang2019-1,Yang2019-2}.

\paragraph{Comparison with experiment and conclusions. ---}
It is instructive to compare our results with experimental data of the LENS group. Although the numerical setup was inspired by the experimental one, our DFT simulations are done for a much smaller system due to high numerical complexity. Also, in calculations we neglect trapping effects along the $z$ direction. Thus, the direct quantitative comparison is limited. Still, we may derive valuable conclusions by performing a qualitative comparison. In general, the experiment admits more dissipative dynamics as observed in the simulations.
As already expected from the static considerations, the temperature effects significantly affect the observed dynamics, see Fig.~\ref{fig:quad}. This points to the crucial role of mutual friction with the normal component. The dissipative mechanism via excitations of the vortex core, while present, emerges to be of secondary importance. 
Including the temperature effects bring us closer to the LENS data. However, even for temperature $T\approx 0.3-0.4\,T_c$ (as reported in the experimental paper), simulations admit weaker dissipation. Note that our BCS runs are done for $a_s\kF=-1$, while in the experiment $a_s\kF=-3.2$, and thus simulations should overestimate the dissipative effects.  The framework applied here currently represents the most complete microscopic description of the fermionic dynamics, without introducing any adjustable (phenomenological) parameters. 
Although the mechanism described by Silaev operates, the lack of two-body collisions is expected to be responsible for effective suppression of dissipation in the theory and deviation from experiment.
In light of these results, we envision that accounting for dissipation and fluctuations by the TDDFT in the future will be inevitable, similar to the case of GPE-like approaches where a certain degree of dissipation, introduced by hand, is presently a common procedure~\cite{doi:10.1146/annurev-conmatphys-031119-050821}. Some works in this direction have already been done~\cite{Bulgac2019}, however, presented ideas need to be validated by experiments. Systematically derived data from vortex collider experiments as a function of temperature and the interaction strength may provide a valuable benchmark for such refinement~\cite{arxiv.2205.04065}. 

\begin{acknowledgments}
The calculations in this Letter were executed by means of the W-SLDA Toolkit~\cite{WSLDAToolkit}. Reproducibility packs are provided in the Supplemental Material~\cite{SM}. They provide complete information needed to reproduce results presented in this Letter. 

We thank W.J. Kwon, G. Roati and K. Xhani for providing experimental data and fruitful discussion and D. P{\k e}cak and M. Tylutki for valuable criticism. 

   This work was supported by the Polish National Science Center (NCN) under Contracts No. UMO-2017/26/E/ST3/00428 (A. Barresi, A. Boulet and G.W.) and No. UMO-2017/27/B/ST2/02792 (P.M.).
  We acknowledge PRACE for awarding us access to resource Piz Daint based in Switzerland at Swiss National Supercomputing Centre (CSCS), Decision No. 2021240031.
  We also acknowledge the Global Scientific Information and Computing Center, Tokyo Institute of Technology for resources at TSUBAME3.0 (project ID: hp210079). 
\end{acknowledgments}

Calculations were executed by A. Barresi, A. Boulet. Data analysis was performed by A. Barresi, A. Boulet and G.W. All authors contributed to discussion and interpretation of the results and to writing of the Letter.

\bibliography{biblio.bib}

\begin{center}
{\bf Supplemental Material for:}\\
{\bf ``Dissipative Dynamics of Quantum Vortices in Fermionic Superfluid''}\\
\end{center}

{\small In this Supplemental Material we provide additional details concerning the processes of vortex imprinting, tracking methods and uncertainty estimation.
}

\setcounter{figure}{2} 
\setcounter{equation}{5} 

\section{Phase imprinting}
The method to generate vortices used in our work relies on the phase properties of the superfluid. The order parameter $\Delta$, by virtue of being a two-dimensional complex quantity, can be written as $\Delta(\bm{r})=|\Delta(\bm{r})|e^{i\phi(\bm{r})}$, where $\phi(\bm{r})$ is its phase. We use 2d geometry and $\bm{r}=(x,y)$. 
In case of a single vortex, located at position $\bm{r}_i$, the phase rotates by $2\pi$ around the point defined by $|\Delta(\bm{r}_i)|=0$. The phase is given by
\begin{equation}
\phi_i(x,y) =\arctan\bigg(\frac{x-x_i}{y-y_i}\bigg).
\end{equation}
We consider the setup consisting of vortex dipole, which are   vortex-antivortex pairs. The phase patterns are assumed to be a superposition of phase patterns of individual vortices: 
	\begin{subequations}
		\label{eq2}
		\begin{align}
			& \phi_{L}(x,y) =\phi_1(x,y)-\phi_2(x,y), \\
			& \phi_{R}(x,y) =\phi_3(x,y)-\phi_4(x,y),
		\end{align}
	\end{subequations}
where $\phi_{L(R)}(x,y)$ is the phase field characterizing the left (right) dipole. The generic structure of the order parameter reads:
	\begin{equation}
		\Delta(x,y)=|\Delta(x,y)|e^{i\phi_{L}(x,y)}e^{i\phi_{R}(x,y)}.
	\end{equation}
	In numerical realization, when searching for static solution, we impose in each iteration the desired phase pattern, while the absolute value of the order parameter is adjusted self-consistently by the computation process. 
	More details can be found on the \href{https://gitlab.fizyka.pw.edu.pl/wtools/wslda/-/wikis/home}{WSLDA Toolkit site}.
The imprinted phase patterns correspond to stationary vortices. Once we start to evolve the solution, the vortices start to move due to their mutual interaction. The sign of each vortex has been chosen in order to have dipoles that move towards the center of the trap. The vortices acquire kinetic energy at the expense of the interaction energy. It is visible as shrinking of the dipole size as they start to accelerate at the beginning of simulation, see also Fig.~\ref{fig:error}. The initial acceleration takes less than $50/\eF$, which is negligible compared to the total simulation time  $t\sim1000/\eF$. The initial part of the trajectory, affected by spurious effect related to imperfection if the imprinting procedure is rejected from further analysis. 
	
\section{Uncertainty estimation of relative distance between vortices}
\begin{figure}[b]
	\includegraphics[width=0.9\linewidth]{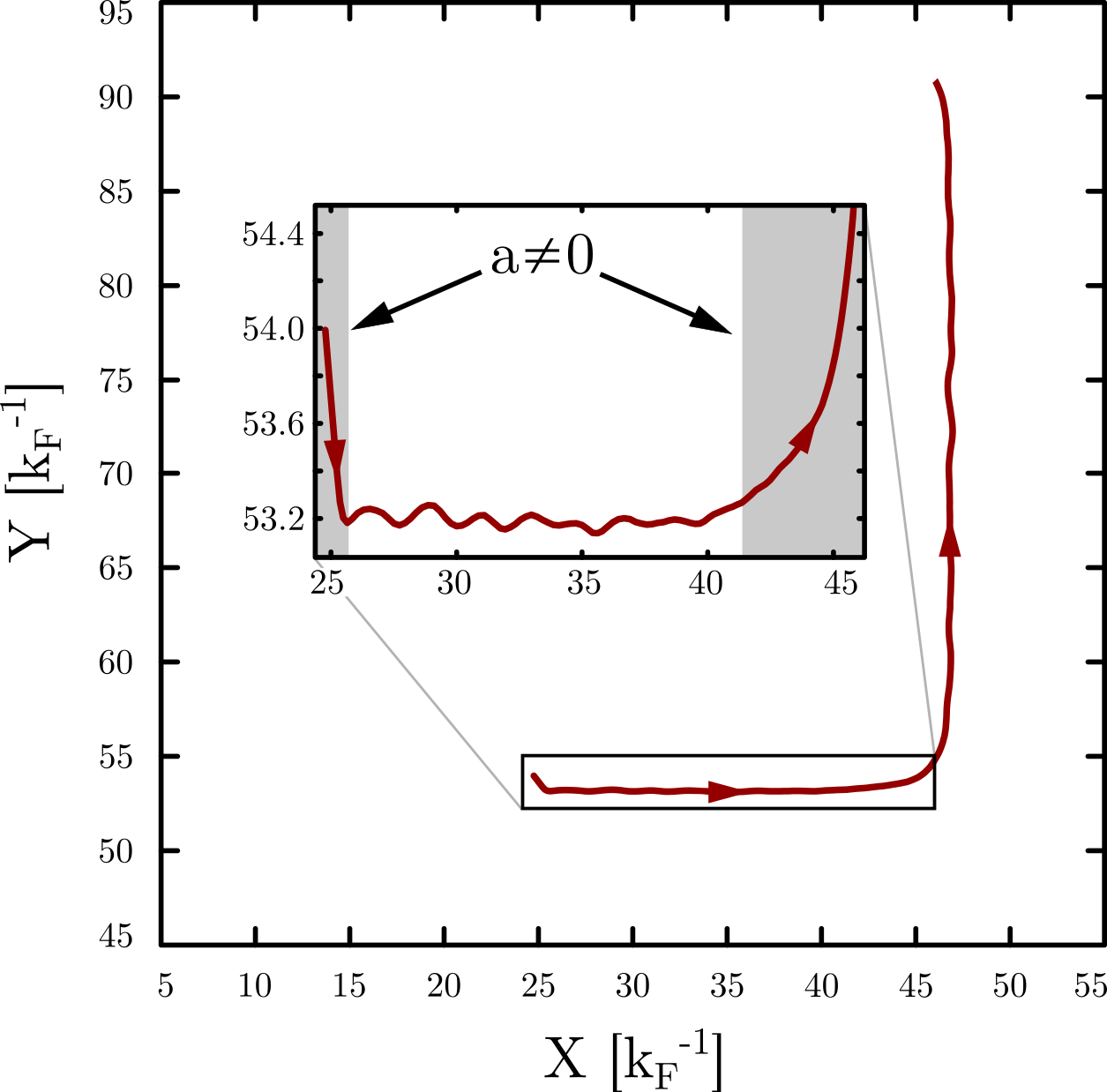}
	\caption[error]{Example trajectory of the top left vortex (positive winding number). Inset: local zoom to show fluctuation of the vortex' trajectory. Greyed out areas have nonzero acceleration, and are excluded from the sample to obtain $d_i$.}
	\label{fig:error}
\end{figure}
To accurately measure the position and the trajectory of each vortex, we use the vortex tracking method as described in Appendix C of Ref.~\cite{PhysRevA.105.013304}. It is a modified version of algorithm presented in~\cite{Villois_2016}, originally constructed  to track vortex lines in superfluids described by the Gross-Pitaevskii equation. The method localizes the vortex core with sublattice resolution, estimated to be about $0.1$ of the lattice spacing. 
\begin{figure}[t]
	\includegraphics[width=0.9\linewidth]{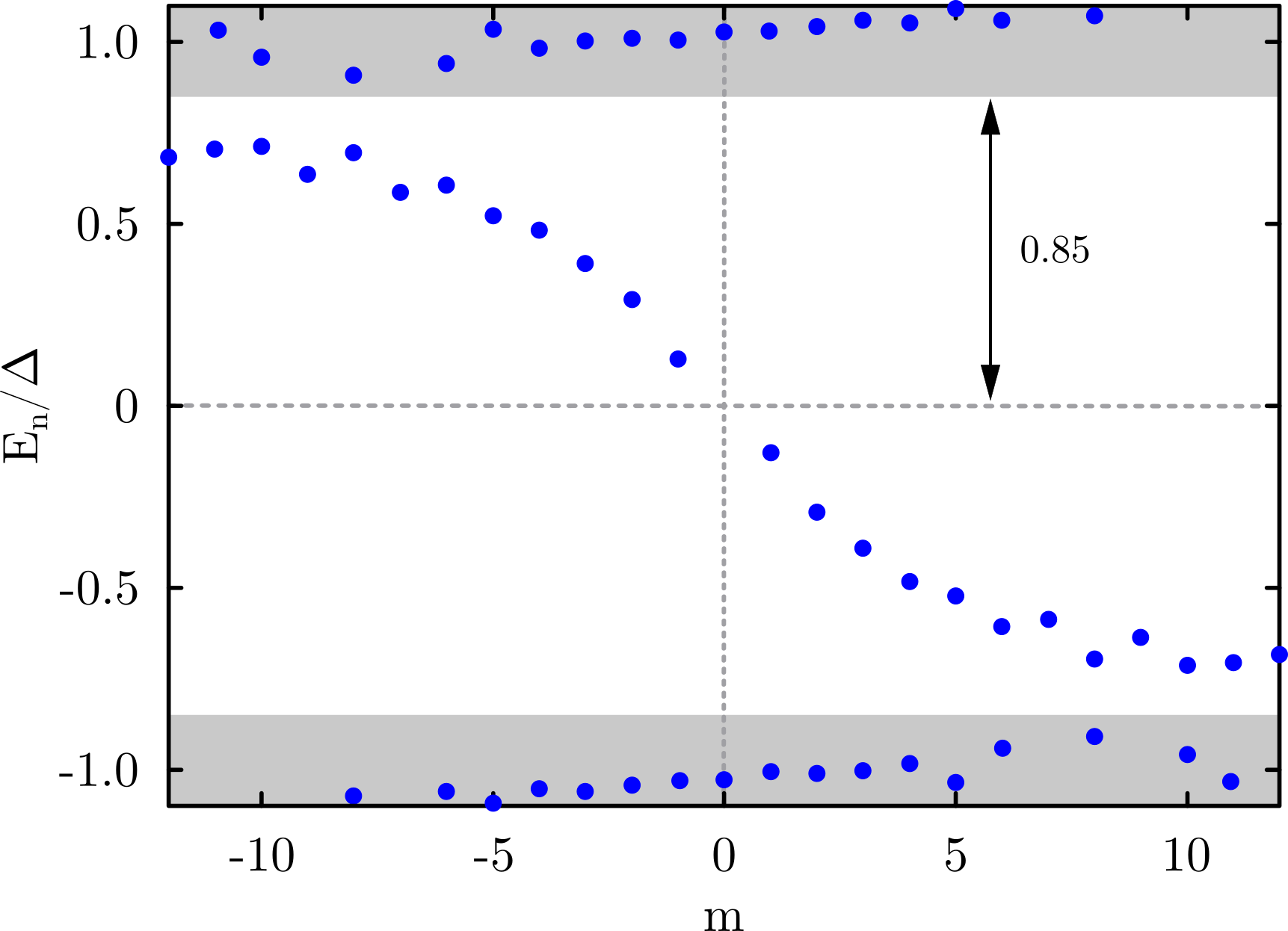}
	\caption[andreev]{Andreev states in the BCS regime as a function of angular momentum quantum number $m$. Symmetry of the spectrum with respect to Fermi surface ($E_n=0$) is due to particle-hole symmetry. The plot is based on data published in paper~\cite{PhysRevA.106.033322}.}
	\label{fig:andreev}
\end{figure}
	
Example of extracted vortex trajectory by the tracking algorithm is presented in Fig.~\ref{fig:error}. When looking closely at the vortex trajectory before (see inset) and after collision they are not perfectly straight lines, but exhibit some small oscillations. In order to estimate an initial distance $\din$ and a final distance $\dout$, we focus on time intervals where vortices move with constant velocity. In practice, this means excluding the time window where the initial acceleration is nonzero (due to the imprinting procedure, see above), and the window around the collision at time $t=t_0$, where the centripetal acceleration becomes nonzero and the trajectories are bent, as shown in Fig. ~\ref{fig:error}.
The size of the fluctuations define uncertainty when computing relative distance $d_{\textrm{i/f}}$ from distance between the two branches, and the uncertainty propagation formula is used when computing the ratio $\dout/\din$ computing.
	
\section{Andreev states tracking}
The Andreev states are defined as states with $|E_n|<\Delta$, where $\Delta$ indicates bulk value of the order parameter. They are discrete and localized states, while states above the gap are delocalized with continuum spectrum. 
However, in numerical realization when we discretize the problem on the lattice, the transition between localized and delocalized states is not sharp. In particular, due to inhomogeneities of the system, some of the states belonging to continuum have energies already slightly below the gap. 
Based on analysis of the spectra $E_n$ we find that (empirical) definition  $|E_n|\lesssim 0.85\Delta$ separates the Andreev states from the continuum states with reasonable accuracy. 
As an example, in Fig.~\ref{fig:andreev}, we show spectra of states for a single vortex in the BCS ($a_s\kF=0.84$) regime, as a function of angular momentum quantum number $m$. The Andreev band (also called chiral band) is clearly visible, and well separated from other states when using the empirical definition. 
This definition is used when computing density arising only from the Andreev states (see Eq.~(4) in the main text). 
\end{document}